\documentclass[preprint,nofootinbib]{revtex4}

\usepackage{dsfont}

\usepackage{graphicx}

\usepackage{amssymb}

\begin{document}

\title{Quasinormal modes and horizon area quantisation in Loop Quantum Gravity}

\author{Saulo Carneiro$^{1,2}$ and C\'assio Pigozzo$^{1}$}

\affiliation{$^{1}$Instituto de F\'isica, Universidade Federal da Bahia, 40210-340, Salvador, BA, Brazil\\ $^{2}$PPGCosmo, CCE, Universidade Federal do Esp\'irito Santo, 29075-910, Vit\'oria, ES, Brazil}

\begin{abstract}
It is argued that the quantum of area between consecutive, high overtones quasinormal modes of a black hole horizon coincides with the area gap predicted by Loop Quantum Gravity, as long as the horizon is isolated and the Barbero-Immirzi parameter is $\gamma \approx \sqrt{3}/6$, in agreement with the value derived from the Bekenstein-Hawking horizon entropy.
\end{abstract}
 
\maketitle

\section{Introduction}
 
The connection between quasinormal modes and quantisation of the horizon area has drawn some attention in the recent past after a pioneering work by Hod \cite{Hod}. On the basis of his approach, the Barbero-Immirzi parameter of Loop Quantum Gravity (LQG) could be fixed \cite{dreyer}, but it was soon shown in disagreement with the value obtained from the entropy of large black holes. In addition, the Hod's main assumption was later criticised by Maggiore \cite{Maggiore}, namely the use of the real part $\omega_R$ of the QNM frequencies in the area quantum estimation. Maggiore considered, instead, a corresponding damped oscillator and used the difference between the natural frequencies of subsequent modes, $\delta \omega_0$. In the high overtones limit of a Schwarzschild black hole of mass $M$, it is possible to show that $\omega_0 \approx \omega_I$ (where $\omega_I$ is the imaginary part of QNM frequencies) and there is a constant shift between consecutive modes, given by $\delta \omega_I = 1/(4 M)$. This led Maggiore to $\delta A = 8\pi l_P^2$, the quantum of area early suggested by Bekenstein on the basis of a different reasoning.\footnote{$l_P^2 = 1$ is the Planck length in natural units $\hbar = c = G =1$.} In the present paper we argue that the correspondence between the horizon area levels and LQG area eigenvalues should only be valid for strictly isolated horizons, for which the Hawking radiation is exactly null. In particular, for the case of a Kerr extremal horizon, Berti et al. \cite{Kerr} numerically showed that $\delta \omega_I = 1/(4 M)$ is still correct within $98\%$ precision. We will show that this allows us to obtain $\delta A = 4\pi l_P^2$, half of the Bekenstein's value.

On the other hand, in Loop Quantum Gravity \cite{Ashtekar,Rovelli,Thiemann,Gambini} the area created when a surface is pierced by a spin network line in the fundamental representation $j =1/2$ is given by $4 \pi \sqrt{3} \gamma l_P^2$ \cite{Smolin}, where $\gamma$ is the Barbero-Immirzi parameter, the only free parameter in the theory. Furthermore, isolated horizons are characterised by the so called projection constraint, which implies that lines with $j = 1/2$ must pierce the horizon at least twice, which leads again to $\delta A = 4\pi l_P^2$, provided that $\gamma = \sqrt{3}/6$. This is in $95\%$ agreement to the approximate value derived from the entropy of large horizons \cite{Meissner}. It also provides, through an exact computation of micro-states of Planck scale horizons, a precise fit to the Bekenstein-Hawking entropy at leading order \cite{Pigozzo}.

\section{Horizon quantisation}

\subsection{Hod's approach}

Black holes linear perturbations vanish in time according to
\begin{equation} \label{QNM}
e^{-\omega_I} [a \sin (\omega_R t) + b \cos (\omega_R t)],
\end{equation}
with a quasinormal frequency 
\begin{equation}
\omega = \omega_R + i\, \omega_I.
\end{equation}
In the case of a Schwarzschild black hole, in the limit of high overtones the quasinormal frequencies are \cite{Zhidenko}
\begin{equation} \label{large_n}
M \omega_n = \frac{\ln 3}{8\pi} - \frac{i}{4} \left( n + \frac{1}{2} \right) + O\left[ (n+1)^{-1/2} \right].
\end{equation}
Hod \cite{Hod} postulated that the energy gap between two subsequent modes is equal to the real part of their quasinormal frequencies, which, in the limit of large $n$, does not depend on $n$, as shown above. Therefore, we have
\begin{equation}
\delta M = \frac{\ln 3}{8\pi M}.
\end{equation}
The area of a Schwarzschild black hole is
\begin{equation} \label{A}
A = 16 \pi M^2,
\end{equation}
leading to
\begin{equation}
\delta A = 32 \pi M \delta M = 4 \ln 3\, l_P^2.
\end{equation}
The area gap in LQG is, in turn,
\begin{equation} \label{LQG}
\delta A = 4\pi \sqrt{3} \gamma l_P^2,
\end{equation}
which gives
\begin{equation}
\gamma = \frac{\ln 3}{\pi \sqrt{3}},
\end{equation}
a value in disagreement with estimates coming from the entropy of large horizons \cite{Meissner} (see, however, \cite{barbero}). Indeed, with the Dogamala-Lewandowski count of horizon micro-states \cite{DL} the theory predicts $\gamma \approx 0.238$, while in the Gosh-Mitra count \cite{GM} we have $\gamma \approx 0.274$.

\subsection{Maggiore's approach}

In Maggiore's approach \cite{Maggiore}, the classical system corresponding to quasinormal modes is a damped harmonic oscillator with natural frequency $\omega_0$. Its solution has the form (\ref{QNM}), with
\begin{equation}
\omega_0^2 = \omega_R^2 + \omega_I^2.
\end{equation}
The energy gap between two subsequent modes is given by the difference between the corresponding values of $\omega_0$. For large $n$, $\omega_I \gg \omega_R$ and we have, from (\ref{large_n}),
\begin{equation}
\delta M \approx \delta \omega_I = \frac{1}{4M}.
\end{equation}
Using (\ref{A}) again, we obtain the quantum of area postulated by Bekenstein,
\begin{equation} \label{bekenstein}
\delta A = 8 \pi l_P^2.
\end{equation}

\subsection{A LQG approach}

In spite of the sub-dominance of the Hawking radiation backreaction when we find exact solutions of large black holes, it is not negligible if our goal is to obtain the area spectrum. In fact, in Maggiore's derivation we have $\delta M = 1/(4M)$, that has the same order of magnitude as the Hawking temperature. Therefore, if we want to identify the horizon area levels as eigenvalues of the LQG area operator, we should consider strictly isolated horizons in order to completely avoid Hawking radiation. Among stationary and asymptotically flat space-times, these are the charged and rotating extremal horizons, with zero surface gravity\footnote{In Refs. \cite{visser,pad} it is argued that the QNM frequencies in the high overtones limit are integer multiples of the surface gravity not only in the Schwarzschild case, but for any spherically symmetric black hole, including the Reissner-Nordstr\"om solution. Their derivations, however, are not valid for extremal horizons.}$^,$\footnote{For a discussion on the quantisation of non-extremal rotating horizons and its possible imprints on gravitational waves from merging black holes, see \cite{agullo,opposite}.}. In the case of an extremal Kerr black hole, the horizon area is given by \cite{BH}
\begin{equation} \label{J}
A = 8 \pi J = 8 \pi M^2,
\end{equation}
from which we have
\begin{eqnarray} \label{13}
\delta A = 16 \pi M \delta M = 16 \pi M \delta \omega_I,
\end{eqnarray}
where we have followed Maggiore in identifying $\delta M$ to $\delta \omega_0$ and taking $\omega_I \gg \omega_R$.
A precise numerical computation of large $n$ quasinormal modes of Kerr black holes \cite{Kerr} has shown that, indeed, $\omega_I \gg \omega_R$ and that
\begin{equation}\label{kerr}
\delta \omega_I \approx 1/2 + 0.0438\, a - 0.0356\, a^2,
\end{equation}
where $a = J/M$ is the black hole angular momentum per unit mass. In the units used in \cite{Kerr}, the extremal case corresponds to $a = M = 1/2$, and we then have $\delta \omega_I \approx 0.513$, $2.6\%$ above the Schwarzschild value $1/(4M)$. Within this precision, Eq. (\ref{13}) can be written
\begin{equation} \label{A''}
\delta A = 4 \pi l_P^2,
\end{equation}
half of the Bekenstein's result. This is, by the way, in accordance to the quantisation of the angular momentum $J$ in integer multiples of $\hbar /2$.

On the other hand, in the realm of Loop Quantum Gravity, horizon isolation corresponds to the projection constraint $\sum_i m_i = 0$, where $m_i \in \{-j_i, -j_i+1,...,j_i-1,j_i\}$, $j_i \in \mathds{Z}_+/2$ are colours of spin network lines piercing the horizon -- irreducible representations of the gauge group {\it SU(2)} -- and the sum is over all the resulting punctures \cite{Perez}. Consider a horizon pierced by a spin network line in the fundamental representation $j = 1/2$. The projection constraint means that the line must pierce the horizon at least twice, generating an area\footnote{A single puncture with $(j,m) = (1,0)$ generates an area smaller than (\ref{A''}), but it does not belong to the area spectrum of an extremal Kerr black hole, as it would add a fraction of $\hbar/2$ to the angular momentum. As a matter of fact, it is easy to see from (\ref{J}) and the area eigenvalues equation that only a spin network of lines with $j = 1/2$ can result in an angular momentum multiple of $\hbar/2$. This means that the LQG area spectrum of large extremal Kerr horizons is equally spaced, as we need in order to fit the QNM spectrum in the high overtone limit.}
\begin{equation}\label{Alqg}
\delta A = 8\pi \gamma l_P^2 \sum_{i=1}^2 \sqrt{j_i(j_i+1)} = 8\pi \sqrt{3} \gamma l_P^2.
\end{equation}
If we identify this area with the quantum given by (\ref{A''}), we obtain $\gamma = \sqrt{3}/6$. This value of the Barbero-Immirzi parameter gives an excellent leading-order fit of the Bekenstein-Hawking entropy of Planck scale black holes \cite{Pigozzo}, and it is only $5\%$ above the approximate value $\approx 0.274$ obtained from the entropy of large horizons in the Gosh-Mitra count of micro-states \cite{Meissner}. The fundamental area of LQG turns out to be $2\pi l_P^2$. When a spin network line with $j=1/2$ crosses the horizon, the angular momentum of an extremal Kerr black hole increases by $\hbar/2$.

\section{Horizon entropy}

\begin{figure}
    \centering
    \includegraphics[scale=.4]{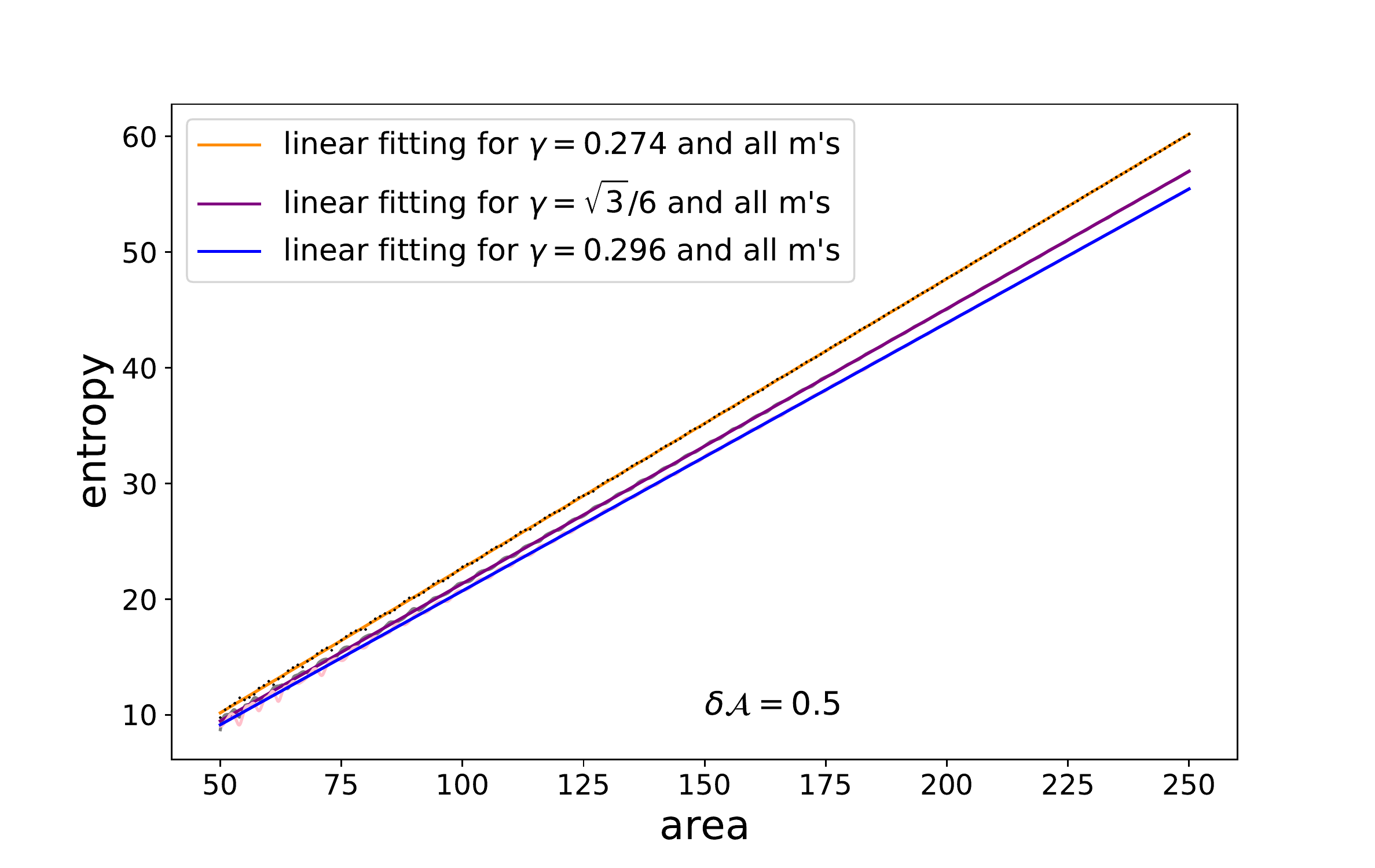}
    \caption{Entropy $\times$ horizon area (in Planck units) for three different values of the Barbero-Immirzi parameter, when the projection constraint is not imposed.}
    \label{fig1}
\end{figure}

\begin{figure}
    \centering
    \includegraphics[scale=.3]{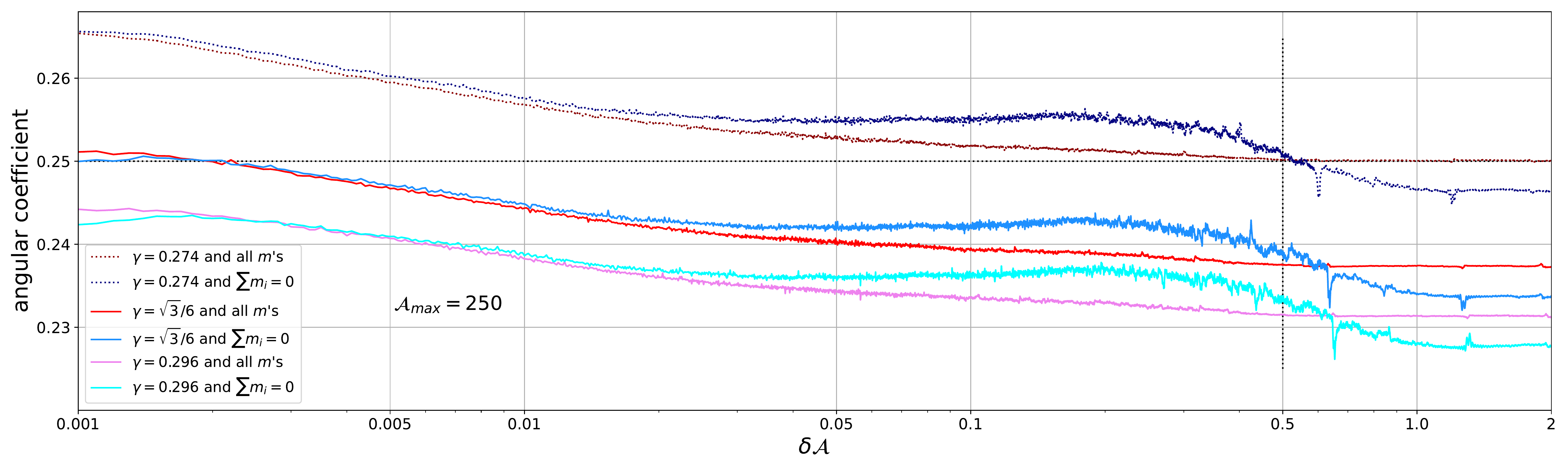}
    \caption{Slope of the entropy $\times$ area leading term as a function of the used area bin, for three different values of the Barbero-Immirzi parameter, with and without the projection constraint.}
    \label{fig2}
\end{figure}

As we have seen, for an extremal rotating black hole the value of $\delta \omega_I$ given by (\ref{kerr}) is $2.6\%$ above $1/(4M)$. Therefore, within the numerical precision of (\ref{kerr}), the value of $\delta {A}$ is higher than  (\ref{A''}) by the same amount and, when comparing with (\ref{Alqg}), we obtain a Barbero-Immirzi parameter $2.6\%$ higher than $\sqrt{3}/6$. As the latter gives a very good fit of the Bekenstein-Hawking entropy \cite{Pigozzo}, we should be concerned about this small difference. We have then redone the analysis of \cite{Pigozzo}, comparing the results derived with three different values of $\gamma$, namely $\gamma = 0.274$ (the approximate value obtained from the entropy of large horizons), $\gamma = \sqrt{3}/6$ and $\gamma = 0.296$ (the precise value got from (\ref{kerr})-(\ref{Alqg})). Since, in the limit of large areas, the slope of the entropy $\times$ area relation is inversely proportional to $\gamma$ \cite{Meissner}, in the latter case we expect a leading order linear relation with slope $\approx 2.6\%$ smaller than $1/4$.

Our computational analysis consists of counting the number of spin network configurations that satisfies the area operator eigenvalues (for details, see \cite{Pigozzo} and references therein). The results depend on the bin used in the variation of $A$, and they are shown in Figs. \ref{fig1} and \ref{fig2} for a maximum area of $250\, l_P^2$. In Fig.~\ref{fig1} we display the entropy $\times$ area relation for the three above values of $\gamma$, without imposing the projection constraint, for a bin $\delta {\cal A} = 0.5$. We see that the oscillations characteristic of small areas are attenuated in the thermodynamic limit of larger areas, as expected \cite{barbero2}. We also see that the slope of the leading linear fit decreases with $\gamma$. In Fig.~\ref{fig2} we show the slope as a function of the are bin, with and without the projection constraint. For large bins the Hawking slope $1/4$ is obtained with $\gamma = 0.274$, as in the limit of large areas \cite{Meissner}. In the limit of small bins, however, the best fit is given by $\gamma = \sqrt{3}/6$. For $\gamma = 0.296$ we have a slope slightly below $1/4$, as expected. The understanding of this small discrepancy between the Barbero-Immirzi parameter derived from quasinormal modes and that obtained from the Bekenstein-Hawking entropy is something deserving further investigation.

\section {Concluding remarks} 

The presence of only one free parameter in the theory turns Loop Quantum Gravity falsifiable if we can find two independent ways of fixing this parameter. Quasinormal modes and horizon thermodynamics are not really independent as both come from the same classical theory. However, the precise coefficient in the linear relation between entropy and area of large horizons has a quantum origin related to
black hole evaporation. In this respect, it seems intriguing that LQG can fix this coefficient with (almost) the same Barbero-Immirzi parameter obtained when we derive the fundamental quantum of area from the Correspondence Principle applied to quasinormal modes. Note that with QNM we only test the ``area gap'' of LQG, while the horizon entropy is a test for the whole area spectrum. None of them can be considered an empirical test, as QNM and the Bekenstein-Hawking entropy are pure theoretical results. Together, they constitute rather a consistency test for the theory, that in this way has no free parameter anymore and can, in principle, be falsified in a single experiment. Another possible signature for the value of $\gamma$ has been recently suggested in \cite{Pigozzo}. There, assuming that quantum corrections to the horizon area are negligible for Planck size horizons\footnote{This assumption is corroborated by effective models of quantum black holes, as e.g. that proposed in \cite{olmedo}. See \cite{Pigozzo} for details.}, it is shown that the horizon area of a Planck mass extremal rotating black hole is in the spectrum of the LQG area operator, provided that $\gamma \approx \sqrt{3}/6$. Incidentally, such stable black holes could be produced at the end of inflation at a rate large enough to be interpreted as composing the presently observed dark matter \cite{PLB}.

\section*{Acknowledgements}

We are thankful to Alberto Saa for helpful discussions and comments. SC is partially support by CNPq (Brazil) with grants \# 311584/2020-9 and 420641/2018-1.

\end{document}